\documentclass[superscriptaddress,showkeys,showpacs,12pt]{revtex4}
\usepackage{graphicx}
\usepackage{amsmath}
\usepackage{booktabs}
\usepackage{amssymb}
\usepackage{multirow}
\usepackage{makecell}
\usepackage{graphicx}
\usepackage{textcomp}
\usepackage{subfigure}
\usepackage{phonetic}
\usepackage{extarrows}
\usepackage{color}
\usepackage{float}
\usepackage[colorlinks,citecolor=blue,linkcolor=blue]{hyperref}
\begin{document}
\title{Optomechanical cooling beyond the quantum backaction limit with frequency modulation}
\author{Dong-Yang Wang}
\affiliation{Department of Physics, Harbin Institute of Technology, Harbin, Heilongjiang 150001, China}
\author{Cheng-Hua Bai}
\affiliation{Department of Physics, Harbin Institute of Technology, Harbin, Heilongjiang 150001, China}
\author{Shutian Liu\footnote{E-mail: stliu@hit.edu.cn}}
\affiliation{Department of Physics, Harbin Institute of Technology, Harbin, Heilongjiang 150001, China}
\author{Shou Zhang\footnote{E-mail: szhang@ybu.edu.cn}}
\affiliation{Department of Physics, College of Science, Yanbian University, Yanji, Jilin 133002, China}
\author{Hong-Fu Wang\footnote{E-mail: hfwang@ybu.edu.cn}}
\affiliation{Department of Physics, College of Science, Yanbian University, Yanji, Jilin 133002, China}

\begin{abstract}
In the usual optomechanical cooling, even if the system has no thermal component, it still has a quantum limit\textemdash known as the quantum backaction limit (QBL)\textemdash on the minimum phonon number related to shot noise. By studying the side-band cooling regime in optomechanical system (OMS), we find that the cooling can be improved significantly when the frequency modulation (FM) that can suppress the Stokes heating processes is introduced into the system. We analyze and demonstrate the reasons of the phonon number below the QBL redefined in the whole stable region of the standard OMSs. The above analyses are further checked by numerically solving the differential equations of second order moments derived from the quantum master equation with broad system parameters, ranging from weak coupling (WC) to ultra-strong coupling (USC) and resolved side-band (RSB) to unresolved side-band (USB) regimes. Comparing with the cases of those without FM, the stable ground-state cooling can also be achieved even in the conventional unstable region.
\end{abstract}
\maketitle

\section{Introduction}\label{sec.1}
Over the past decades, special attention has been paid to studying the micromechanical resonators which are novel quantum devices and are used to explore various of quantum mechanics questions of the macroscopic scale, such as quantum-classical mechanics boundary~\cite{PhysRevLett.88.120401,PhysRevLett.97.237201,POOT2012273}, ultrahigh precision metrology~\cite{RevModPhys.52.341,RevModPhys.68.755}, and gravitational-wave detection~\cite{PhysRevLett.116.061102}, etc. However, for overcoming thermal noise to achieve those applications, the mechanical resonator needs to be cooled to its ground state at first. So far, various of proposals of cooling resonator have been proposed and discussed, including feedback control~\cite{PhysRevLett.80.688,Wilson2015}, modulating coupling strength~\cite{PhysRevLett.95.097204,PhysRevLett.107.177204}, and cavity side-band cooling~\cite{PhysRevLett.99.093901,PhysRevLett.99.093902}. In these proposals mentioned above, cavity side-band cooling method is one of the most promising proposals, which has been studied widely and realized experimentally~\cite{nature.444.67,Nat.Phys.4.415,Nature.463.72,1367-2630-14-9-095015,PhysRevLett.116.013602}. However, there has a QBL on the minimum phonon number related to shot noise even if the systems have no any coupling with thermal environment~\cite{PhysRevLett.99.093902,CPB.22.114213}. Recently, the QBL of cooling mechanical resonator has also been reached in cavity OMS via the side-band cooling approach~\cite{PhysRevLett.116.063601}. It is natural to think that the cooling limit can be broken through some ways. 

To this end, some theoretical proposals have been proposed to cool the mechanical resonator below the QBL, including dissipative coupling~\cite{PhysRevLett.102.207209}, pulsed cooling~\cite{PhysRevLett.110.153606}, electromagnetically-induced-transparency-like~\cite{PhysRevA.90.013824,PhysRevA.90.053841,Liu:18,PhysRevA.93.033845}, trapping optical parametric amplifier~\cite{PhysRevA.79.013821}, and injecting squeezed light~\cite{PhysRevA.94.051801}, etc. And the proposal utilizing squeezed light to improve cooling has been proved experimentally~\cite{nature.541.191}. Moreover, another interesting experiment also cools the mechanical resonator below the QBL via feedback-controllably engineering the pump field~\cite{PhysRevLett.119.123603}. On the other hand, the modulated OMSs exhibit rich behaviors and have attracted amount of attention~\cite{PhysRevLett.103.213603,PhysRevA.83.043804,PhysRevA.86.013820,PhysRevA.93.033853,PhysRevA.89.023843,PhysRevA.92.013822,PhysRevA.94.053807,PhysRevA.95.053861}. However, in those modulation proposals, the studying about mechanical resonator cooling is relatively rare. In Ref.~\cite{PhysRevA.91.023818}, the authors study the optomechanical cooling theoretically using an adiabatic approximation when the frequency and damping of mechanical resonator are modulated periodically. As far as we know, the approach of directly using FM to improve mechanical cooling below the QBL has not yet been reported. 

In this paper, we propose a proposal to improve micromechanical resonator cooling in OMS via modulating frequencies of both the optical and mechanical components. The FM of optical component is easy to implement and the modulation of micromechanical resonators has also been reported~\cite{PhysRevA.91.023818,JAP.1.2785018,nnano.2013.232,s151026478,nnano.2014.168,nl500879k,1367-2630-9-2-035}. Here we provide a complete and simple understanding of the physical processes about improving mechanical cooling, which allows us to illustrate the deep reasons of the lower mechanical cooling. We consider a conventional OMS in which the mechanical resonator is coupled to optical cavity field via radiation pressure, and the cavity and mechanical resonator are modulated periodically. In such an OMS, we show that the Stokes heating processes can be fully suppressed via FM, and the physical mechanism is also explained and demonstrated through the Raman-scattering and frequency domain pictures. In addition, since the conventional QBL defined as in Ref.~\cite{RevModPhys.86.1391} is satisfied only in the WC region, we recalculate the QBL and steady-state final mean phonon number (MPN) of the standard OMSs in the absence of FM. Moreover, the dynamical evolution of MPN with FM is also obtained by numerically solving the differential equations of second order moments, which are derived strictly from the quantum master equation. And the results are shown respectively with very broad system parameters, e.g., the optomechanical coupling strength ranging from WC to USC and cavity decay rate ranging from RSB to USB regimes. As we all know, the ground-state cooling cannot be achieved when either the coupling strength or cavity decay rate is too large in the standard OMSs without FM. However, the stable lower MPN can be achieved in the presence of FM even in the conventional unstable region of the standard OMSs. Different from the previous methods in Refs.~\cite{PhysRevA.90.013824,PhysRevA.90.053841,Liu:18,PhysRevA.93.033845,PhysRevA.79.013821,PhysRevA.94.051801}, the mechanical resonator cooling is achieved below QBL only through the FM, which does not need extra technologies, such as trapping optical parametric amplifier, injecting squeezed light, or transparent window induced by auxiliary qubit, etc.

The paper is organized as follows: In Sec.~\ref{sec.2}, we derive the linearized Hamiltonian of the OMS with synchronous FM and explain the physical mechanism of suppressing the Stokes processes through the Raman-scattering and frequency domain pictures. In Sec.~\ref{sec.3}, firstly, we give the analytical expression of the steady-state final MPN in the absence of FM and discuss the QBL in the whole stable region. Then we study the dynamical evolution of MPN with broad system parameters through solving numerically the differential equations of second order moments derived from the quantum master equation. Lastly, a conclusion is given in Sec.~\ref{sec.4}.

\section{System and Hamiltonian}\label{sec.2}
We consider a generic OMS in which the mechanical resonator is coupled to a driven optical cavity via radiation pressure. The simplest possible cavity OMS model has been used to describe successfully most of the experiments to date. In the rotating frame at the driven laser frequency $\omega_{l}$, the Hamiltonian of the system is written as
\begin{eqnarray}\label{e01}
H=\Delta_{c}a^{\dagger}a+\omega_{m}b^{\dagger}b-ga^{\dagger}a(b^{\dagger}+b)+(Ea^{\dagger}+E^{\ast}a),
\end{eqnarray}
where $a~(b)$ and $a^{\dagger}~(b^{\dagger})$ represent the annihilation and creation operators for optical (mechanical) mode with frequency $\omega_{c}~(\omega_{m})$, respectively. $\Delta_{c}=\omega_{c}-\omega_{l}$ is the cavity laser detuning, $g$ is the single photon optomechanical coupling strength~\cite{PhysRevApplied.9.064006}, and $E$ is the driving amplitude of laser. Following the usual linearization approach, 
$a=\langle a\rangle+\delta a=\alpha+\delta a$ and $b=\langle b\rangle+\delta b=\beta+\delta b$, the linearized Hamiltonian reads
\begin{eqnarray}\label{e02}
H_{L}=\Delta_{c}^{'}\delta a^{\dagger}\delta a+\omega_{m}\delta b^{\dagger}\delta b
	-(G\delta a^{\dagger}+G^{\ast}\delta a)(\delta b^{\dagger}+\delta b),
\end{eqnarray}
where $\Delta_{c}^{'}=\Delta_{c}-g(\beta+\beta^{\ast})$ and $G=g\alpha$ is linearized optomechanical coupling strength. 
The Hamiltonian in Eq.~(\ref{e02}) corresponds to the standard quantum Rabi model~\cite{PhysRevA.97.033807} including the beam-splitter interaction 
$(G\delta a^{\dagger}\delta b+G^{\ast}\delta a\delta b^{\dagger})$ and 
two-mode squeezing interaction $(G\delta a^{\dagger}\delta b^{\dagger}+G^{\ast}\delta a\delta b)$. In the usual side-band cooling mechanism, the resonant conditions of the beam-splitter interaction are an essential prerequisite to enhance (suppress) the anti-Stokes (Stokes) process, where the strict restrictions of small cavity decay and weak optomechanical coupling strength need to be satisfied. To break those restrictions, we introduce a cosine modulation of the free terms into the initial Hamiltonian. For simplicity, the modulation Hamiltonian is given by
\begin{eqnarray}\label{e03}
H_{M}=\frac{1}{2}\xi\nu\cos(\nu t)(a^{\dagger}a+b^{\dagger}b),
\end{eqnarray}
where $\xi$ is the normalized modulation amplitude and $\nu$ is the modulation frequency. In the superconducting OMSs, the modulation of optical mode can be realized easily by tuning the magnetic flux in superconducting systems. Meanwhile, the modulation of mechanical mode can be realized by tuning the voltage of the gate electrode for graphene resonator~\cite{nnano.2013.232,nnano.2014.168,nl500879k}, which can also be coupled superconducting qubit. In the presence of FM, we can derive the linearized Hamiltonian using the usual linearization approach (see Appendix~\ref{App1})
\begin{eqnarray}\label{e04}
H_{\mathrm{LM}}=\Delta_{c}^{'}\delta a^{\dagger}\delta a+\omega_{m}\delta b^{\dagger}\delta b-(G\delta a^{\dagger}+G^{\ast}\delta a)
	(\delta b^{\dagger}+\delta b)+\frac{1}{2}\xi\nu\cos(\nu t)(\delta a^{\dagger}\delta a+\delta b^{\dagger}\delta b).
\end{eqnarray}

\begin{figure}
	\centering
	\includegraphics[width=0.6\linewidth]{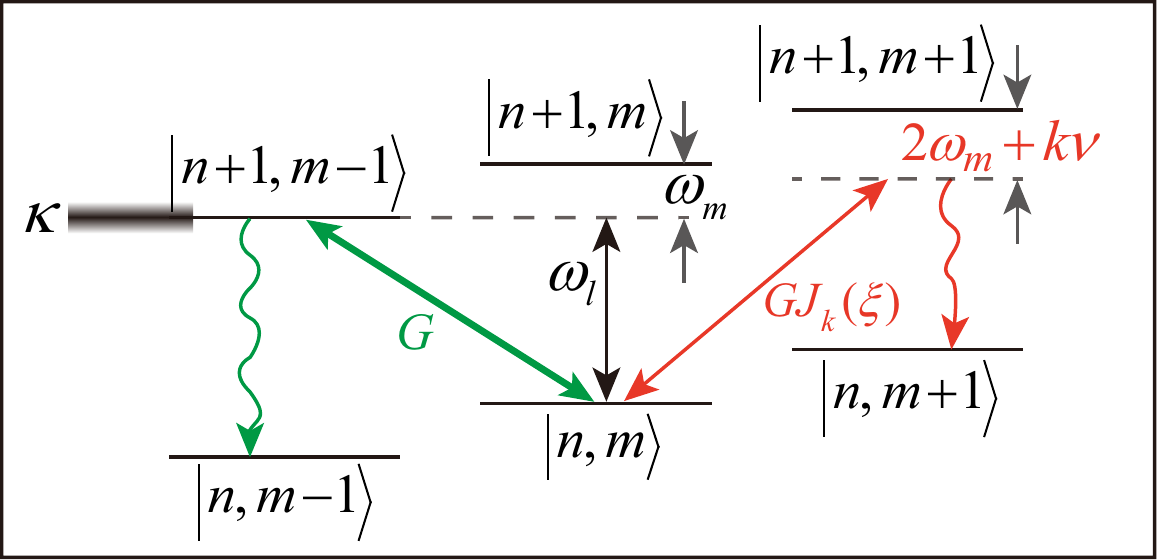}
	\caption{(Color online) Level diagram of the linearized Hamiltonian with FM (Eq.~\ref{e07}). $|n, m\rangle$ denotes the state of $n$ photons and $m$ phonons in the displaced frame. The green (red) arrows represent the cooling (heating) corresponding to the anti-Stokes (Stokes) process.}
	\label{fig:raman-scattering}
\end{figure}

To study the effect of FM on the system dynamics clearly, it is useful to perform the rotating transformation defined by
\begin{eqnarray}\label{e05}
V_{2}&=&\mathcal{T}\exp\left\{-i\int_{0}^{t}d\tau\left[\Delta_{c}^{'}\delta a^{\dagger}\delta a+\omega_{m}\delta b^{\dagger}\delta b+\frac{1}{2}\xi\nu\cos(\nu\tau)(\delta a^{\dagger}\delta a+\delta b^{\dagger}\delta b)\right]\right\}\cr\cr
&=&\exp\left[-i\Delta_{c}^{'}t\delta a^{\dagger}\delta a-i\omega_{m}t\delta b^{\dagger}\delta b-\frac{i}{2}\xi\sin(\nu t)(\delta a^{\dagger}\delta a+\delta b^{\dagger}\delta b)\right],
\end{eqnarray}
where $\mathcal{T}$ denotes the time ordering operator. In the rotating frame defined by the transformation operator $V_{2}$, the transformed Hamiltonian becomes
\begin{eqnarray}\label{e06}
\widetilde{H}_{\mathrm{LM}}&=&V_{2}^{\dag}H_{\mathrm{LM}}V_{2}-iV_{2}^{\dag}\dot{V}_{2}\cr\cr
				&=&-G\left(\delta a^{\dagger}\delta b^{\dagger}e^{i[(\Delta_{c}^{'}+\omega_{m})t+\xi\sin(\nu t)]}
				+\delta a^{\dagger}\delta be^{i(\Delta_{c}^{'}-\omega_{m})t}\right)+\mathrm{H.c.}.
\end{eqnarray}
Using the Jacobi-Anger expansions: $e^{i\xi\sin(\nu t)}=\sum_{k=-\infty}^{\infty}J_{k}(\xi)e^{ik\nu t}$, the transformed Hamiltonian in Eq.~(\ref{e06}) can be rewritten as
\begin{eqnarray}\label{e07}
\widetilde{H}_{\mathrm{LM}}=-G\delta a^{\dagger}\delta be^{i(\Delta_{c}^{'}-\omega_{m})t}
-\sum_{k=-\infty}^{\infty}GJ_{k}(\xi)\delta a^{\dagger}\delta b^{\dagger}e^{i(\Delta_{c}^{'}+\omega_{m}+k\nu)t}+\mathrm{H.c.},
\end{eqnarray}
where $J_{k}(\xi)$ is the Bessel function of the first kind with $k$ being an integer. It is worth noting that, except for the usual linearized approach, our calculations have not taken any other approximations. The further understanding of the physics process can be explained in a Raman-scattering picture, as shown in Fig.~\ref{fig:raman-scattering}, where $|n, m\rangle$ denotes the state of $n$ photons and $m$ phonons in the displaced frame. Under the red detuning side-band resonant condition $(\Delta_{c}^{'}=\omega_{m})$, the anti-Stokes process (the green arrows) is on resonance leading to enhance for the mechanical cooling. Different from the anti-Stokes process, the Stokes processes (the red arrows) exist the detuning $(2\omega_{m}+k\nu)$ with coupling strength $GJ_{k}(\xi)$, which are able to be modulated independently by choosing appropriate parameters $\xi$ and $\nu$. Naturally, if the ratio $GJ_{k}(\xi)/(2\omega_{m}+k\nu)$ can be reduced significantly by modulating those two parameters, the heating (Stokes) process will be suppressed efficiently. However, in the absence of FM, the ratio is a constant $G/(2\omega_{m})$, which implies that the efficient mechanical motion cooling needs to satisfy the restriction condition $G\ll2\omega_{m}$ to suppress the Stokes heating process.

\begin{figure}
	\centering
	\includegraphics[width=0.6\linewidth]{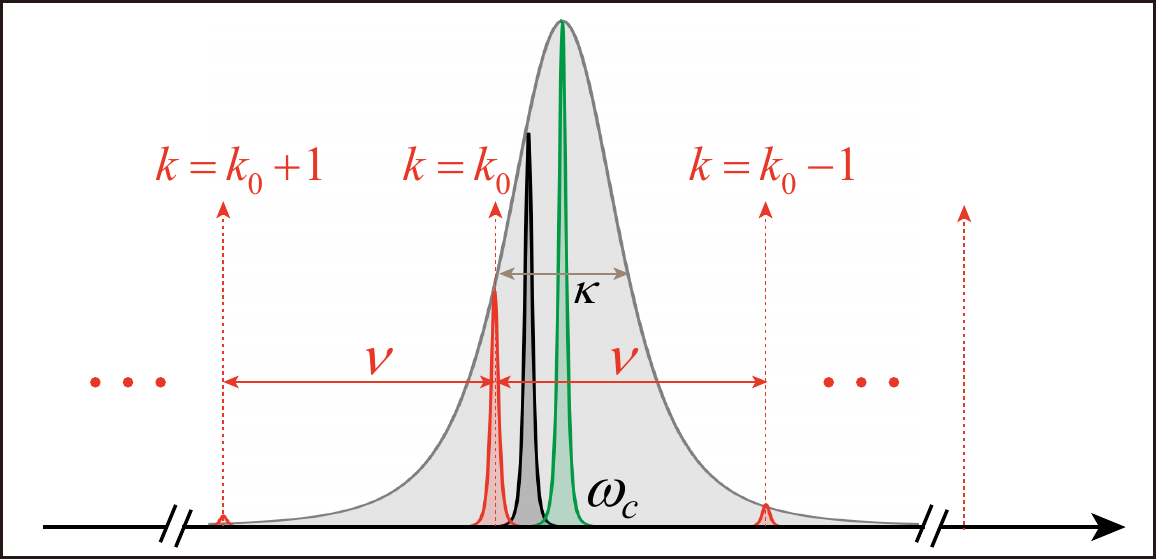}
	\caption{(Color online) Frequency domain interpretation of the optomechanical interactions in the presence of FM.}
	\label{fig:Red-side-band}
\end{figure}

\begin{figure}
	\centering
	\includegraphics[width=0.6\linewidth]{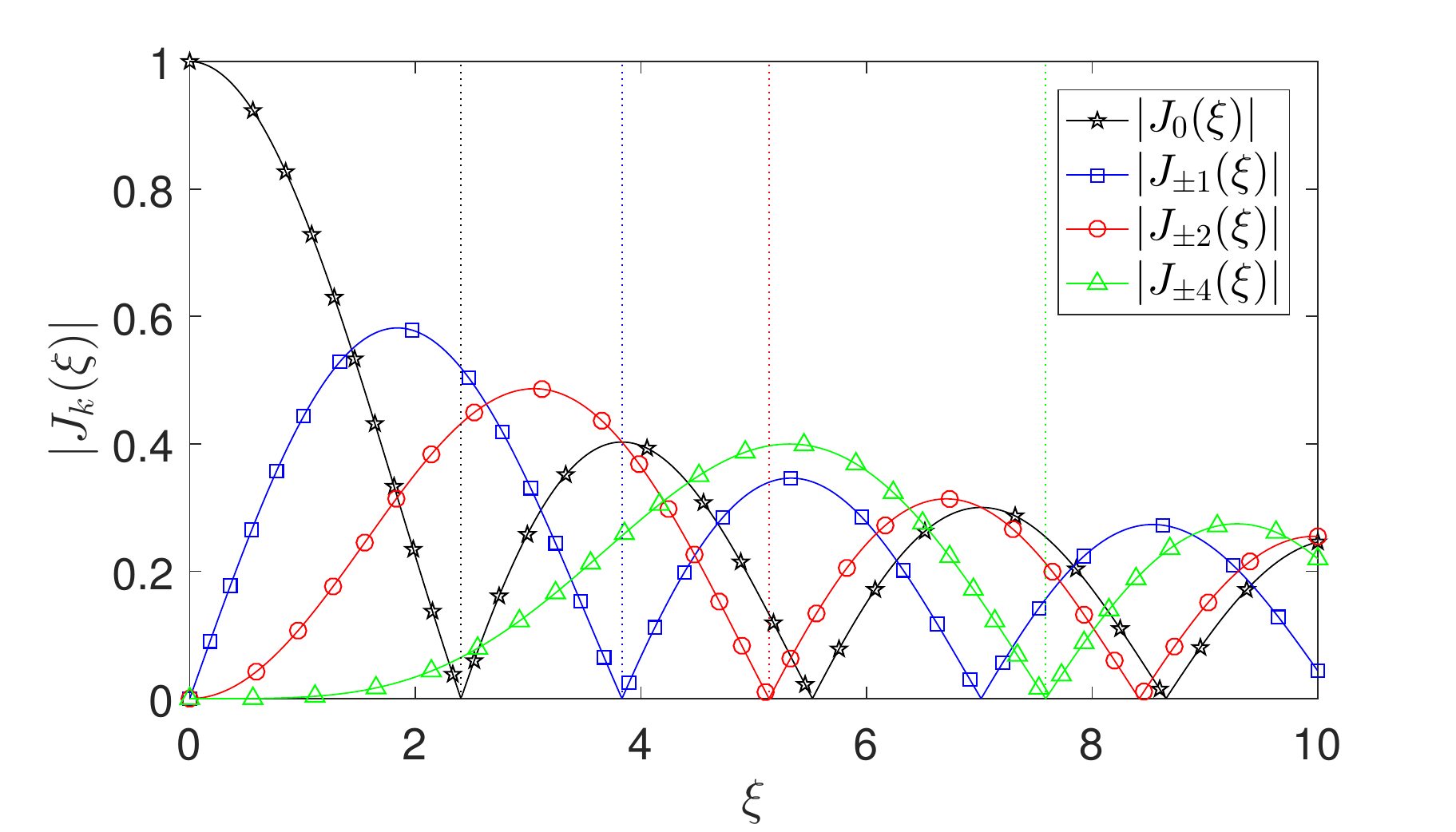}
	\caption{(Color online) The absolute value of the Bessel function of the first kind $|J_{k}(\xi)|$ versus the variables $\xi$.}
	\label{fig:Besselfun}
\end{figure}

Furthermore, the physical mechanism of improving cooling can also be understood by using the frequency domain interpretation, as shown in Fig.~\ref{fig:Red-side-band}, where the Stokes heating processes (the red peaks) are discrete and have been separated with modulation frequency $\nu$ space. For a given value of $\nu$, there exists a corresponding $k=k_{0}$ to make sure the detuning $(2\omega_{m}+k_{0}\nu)$ minimum in all side-bands of Stokes. If the modulation frequency $\nu$ is large enough, the other Stokes processes will be far away the resonant condition and the corresponding photons also cannot exist in the cavity, which lead to those contributions to heating mechanical resonator negligible. The primary purpose is to adopt an appropriate $\xi$ to reduce the ratio $GJ_{k_{0}}(\xi)/(2\omega_{m}+k_{0}\nu)$ for suppressing the nearest resonant heating process, which is the major obstacle for cooling the mechanical resonator. That is easy to achieve because there are always a series of parameters $\xi$ to satisfy $J_{k_{0}}(\xi)=0$, as shown in Fig.~\ref{fig:Besselfun}. Therefore, the nearest resonant heating process will also be negligible. Different from the usual side-band cooling proposals, which rely on the RSB regime $\kappa<\omega_{m}$, the method we proposed can suppress the nearest resonant heating process efficiently through manipulating its coupling strength $GJ_{k_{0}}(\xi)$ even in USB regime, namely, even if the photons of resonating to Stokes processes can exist abundantly in the cavity, the photons cannot interact with the mechanical resonator.

Based on the above analyses, all the Stokes heating processes can be suppressed via FM approach, which strongly indicates that the QBL of mechanical cooling can be broken in the presence of FM. Next, we will verify that how it can be broken via calculating the MPN.

\section{Improving mechanical cooling via frequency modulation}\label{sec.3}
In this section, we study the dynamical evolution behavior of the phonon number under the linearized Hamiltonian in Eq.~(\ref{e04}). It is not necessary to calculate the whole density matrix $\rho$ to obtain the dynamical evolution of MPN. Here, we just need to solve a set of differential equations about the mean values of all the second order moments $\langle\delta a^{\dagger}\delta a\rangle$, $\langle\delta b^{\dagger}\delta b\rangle$, $\langle\delta a^{\dagger}\delta b\rangle$, $\langle\delta a\delta b\rangle$, $\langle\delta a^{2}\rangle$, and $\langle\delta b^{2}\rangle$, which can be derived via the quantum master equation
\begin{eqnarray}\label{e08}
\dot{\rho}=-i\left[H_{\mathrm{LM}},\rho\right]+\kappa\mathcal{L}[a]\rho+\gamma\left(n_{\mathrm{th}}+1\right)\mathcal{L}[b]\rho
+\gamma n_{\mathrm{th}}\mathcal{L}[b^{\dag}]\rho,
\end{eqnarray}
where $\mathcal{L}[o]\rho=o\rho o^{\dag}-(o^{\dag}o\rho+\rho o^{\dag}o)/2$ is the standard Lindblad operators. $\kappa$ and $\gamma$ are the decay rate of optical cavity and the damping rate of mechanical resonator, respectively. Through the quantum master equation in Eq.~(\ref{e08}) and $\mathrm{d}\langle o_{i}o_{j}\rangle/\mathrm{d}t=\mathrm{Tr}(\dot{\rho}o_{i}o_{j})$, where $o_{i}$ and $o_{j}$ are the arbitrary operators of the system, the derivatives of those second order moments are given as~\cite{PhysRevLett.110.153606,NJP.10.095007}
\begin{eqnarray}\label{e09}
\frac{\mathrm{d}}{\mathrm{d}t}\langle\delta a^{\dagger}\delta a\rangle&=&i\left(G\langle\delta a^{\dagger}\delta b\rangle
		-G^{\ast}\langle\delta a^{\dagger}\delta b\rangle^{\ast}+G\langle\delta a\delta b\rangle^{\ast}
		-G^{\ast}\langle\delta a\delta b\rangle\right)-\kappa\langle\delta a^{\dagger}\delta a\rangle,\cr\cr
\frac{\mathrm{d}}{\mathrm{d}t}\langle\delta b^{\dagger}\delta b\rangle&=&i\left(-G\langle\delta a^{\dagger}\delta b\rangle
		+G^{\ast}\langle\delta a^{\dagger}\delta b\rangle^{\ast}+G\langle\delta a\delta b\rangle^{\ast}
		-G^{\ast}\langle\delta a\delta b\rangle\right)-\gamma\langle\delta b^{\dagger}\delta b\rangle+\gamma n_{\mathrm{th}},\cr\cr
\frac{\mathrm{d}}{\mathrm{d}t}\langle\delta a^{\dagger}\delta b\rangle&=&
		\left[i\left(\Delta-\Omega_{m}\right)-\frac{\kappa+\gamma}{2}\right]\langle\delta a^{\dagger}\delta b\rangle
		+i\left(G^{\ast}\langle\delta a^{\dagger}\delta a\rangle-G^{\ast}\langle\delta b^{\dagger}\delta b\rangle
		+G\langle\delta a^{2}\rangle^{\ast}-G^{\ast}\langle\delta b^{2}\rangle\right),\cr\cr
\frac{\mathrm{d}}{\mathrm{d}t}\langle\delta a\delta b\rangle&=&
		-\left[i\left(\Delta+\Omega_{m}\right)+\frac{\kappa+\gamma}{2}\right]\langle\delta a\delta b\rangle
		+i\left(G\langle\delta a^{\dagger}\delta a\rangle+G\langle\delta b^{\dagger}\delta b\rangle+G
		+G^{\ast}\langle\delta a^{2}\rangle+G\langle\delta b^{2}\rangle\right),\cr\cr
\frac{\mathrm{d}}{\mathrm{d}t}\langle\delta a^{2}\rangle&=&
		-\left(2i\Delta+\kappa\right)\langle\delta a^{2}\rangle+2i\left(G\langle\delta a^{\dagger}\delta b\rangle^{\ast}
		+G\langle\delta a\delta b\rangle\right),\cr\cr
\frac{\mathrm{d}}{\mathrm{d}t}\langle\delta b^{2}\rangle&=&
		-\left(2i\Omega_{m}+\gamma\right)\langle\delta b^{2}\rangle+2i\left(G\langle\delta a^{\dagger}\delta b\rangle
		+G^{\ast}\langle\delta a\delta b\rangle\right),
\end{eqnarray}
where $\Delta=\Delta_{c}^{'}+\frac{1}{2}\xi\nu\cos(\nu t)$ and $\Omega_{m}=\omega_{m}+\frac{1}{2}\xi\nu\cos(\nu t)$ are the renormalized parameters. It is worth noting that the dynamical evolution of MPN is exact via solving the differential equations and the dynamics dimension is also not limited as the original quantum master equation. The system stability can also be determined by observing the dynamical behaviors of MPN in the presence of FM.

\subsection{Cooling Limits in the absence of frequency modulation}
Firstly, as a comparison, we give the quantum cooling limit for the whole stable region in the absence of FM $(\xi=0)$, which can be derived from Eq.~(\ref{e09}) when the system finally reaches the steady state, namely, $\mathrm{d}\langle o_{i}o_{j}\rangle/\mathrm{d}t=0$. Under the red side-band resonant condition $(\Delta_{c}^{'}=\omega_{m})$ and the cooperativity parameter $C\equiv4|G|^{2}/(\gamma\kappa)\gg1$, the steady-state final MPN is
\begin{eqnarray}\label{e010}
	\langle\delta b^{\dagger}\delta b\rangle_{\mathrm{lim}}\simeq\frac{4|G|^{2}+\kappa^{2}}{4|G|^{2}(\kappa+\gamma)}\gamma n_{\mathrm{th}}
	+\frac{(4\omega_{m}^{2}-\kappa^{2})(8|G|^{2}+\kappa^{2})+2\kappa^{4}}{16\omega_{m}^{2}(4\omega_{m}^{2}
		+\kappa^{2}-16|G|^{2})},
\end{eqnarray}
where the first term represents the classical cooling limit corresponding to the thermal environment, while the second term represents the quantum cooling limit, as the redefined QBL, corresponding to the cavity decay and quantum backaction. Under the red side-band resonant condition, the conventional steady-state final MPN is defined as (see Appendix~\ref{App2})

\begin{figure}
	\centering
	\includegraphics[width=0.6\linewidth]{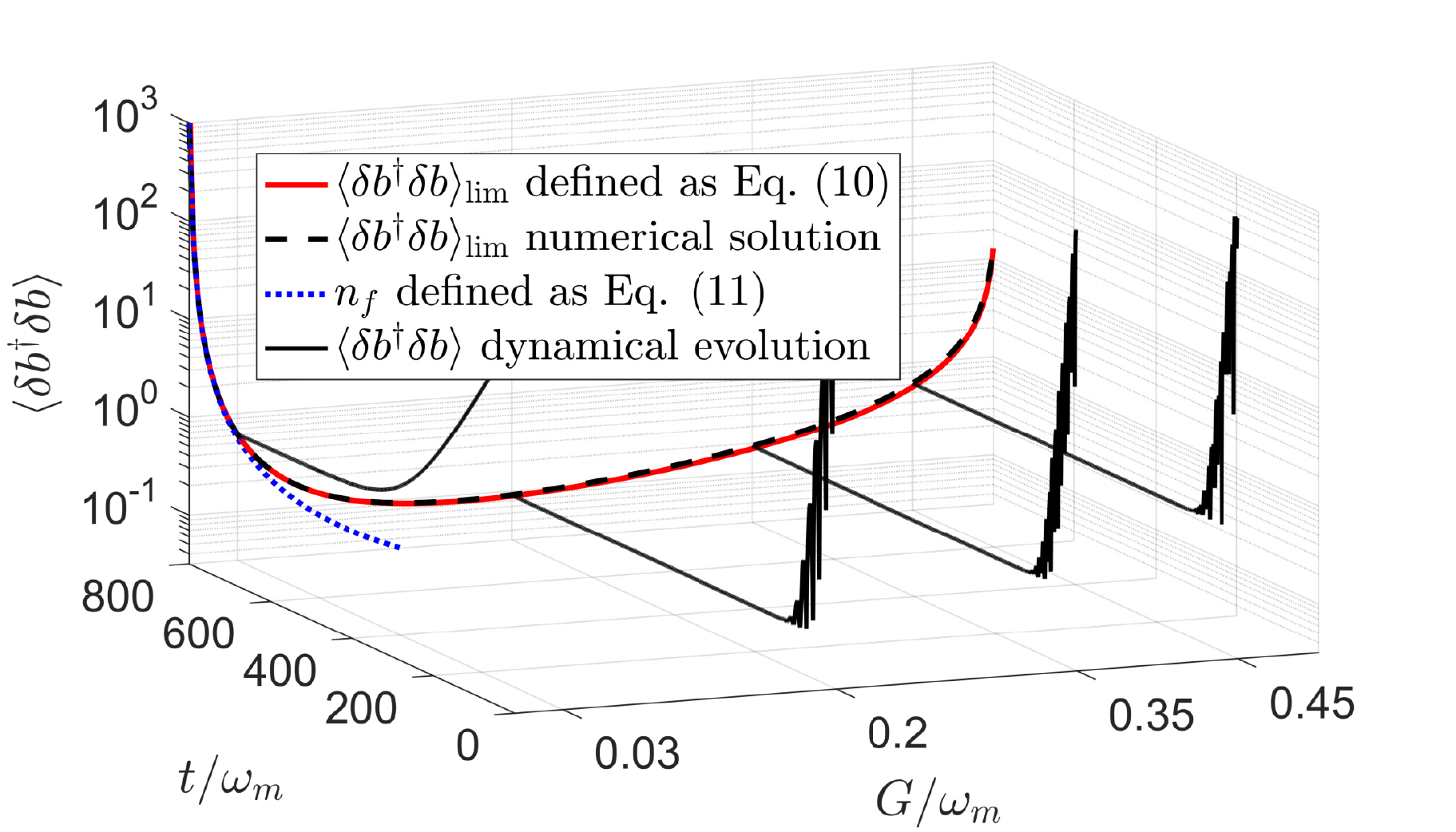}
	\caption{(Color online) The cooling limit varies with the optomechanical coupling strength $G/\omega_{m}$ (red solid line, black dashed line, and blue dot line). The black solid lines represent the dynamical evolution of MPN for different optomechanical coupling strengths in the absence of FM. The other parameters are set as $\Delta_{c}^{'}=\omega_{m}$, $\gamma=10^{-5}\omega_{m}$, $n_{\mathrm{th}}=10^{3}$, and $\kappa=0.2\omega_{m}$.}
	\label{fig:phonon_num_lim_G}
\end{figure}
\begin{figure}
	\centering
	\includegraphics[width=0.6\linewidth]{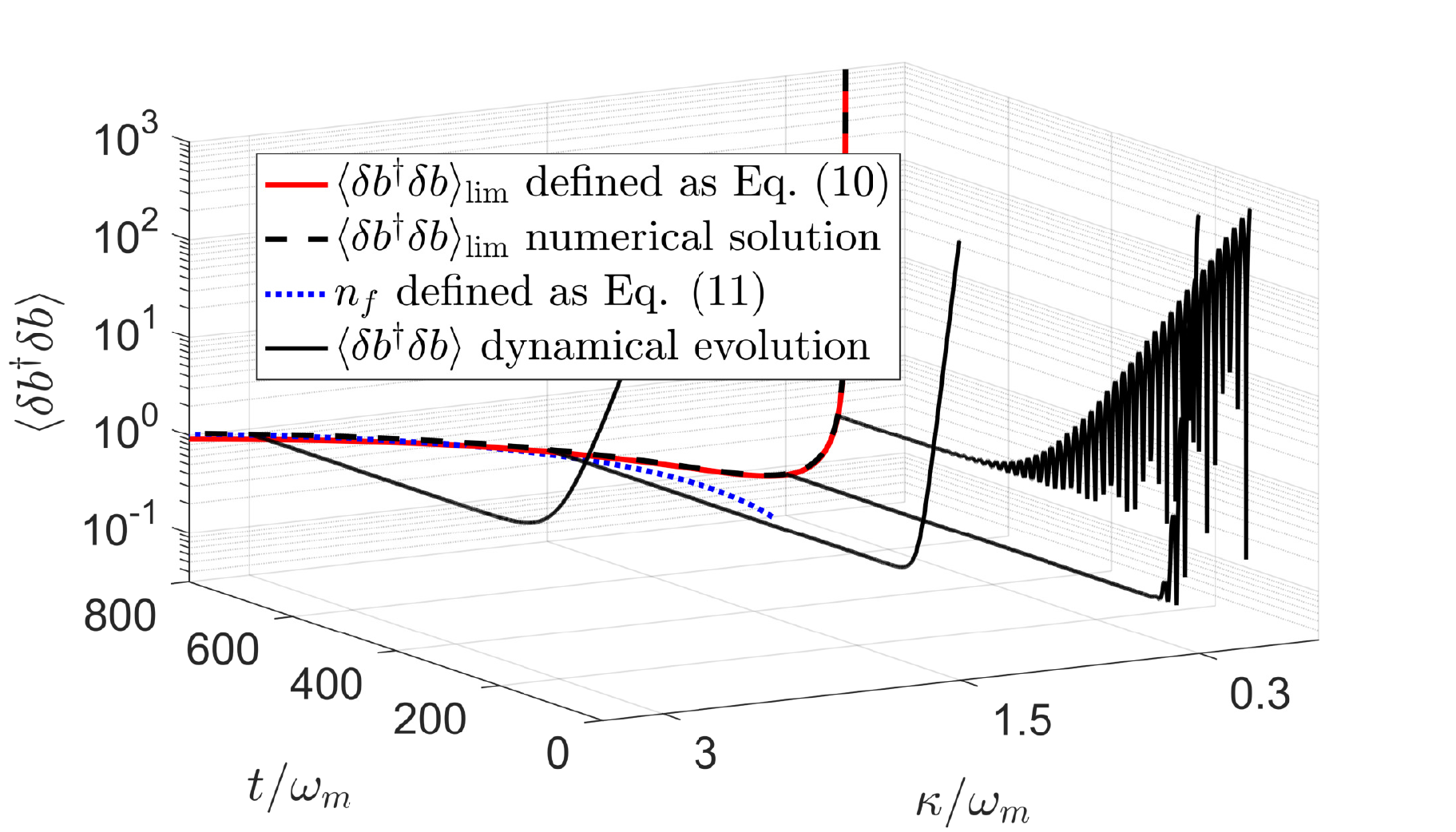}
	\caption{(Color online) The cooling limit varies with cavity decay rate $\kappa/\omega_{m}$ (red solid line, black dashed line, and blue dot line). The black solid lines represent the dynamical evolution of MPN for different cavity decay rates in the absence of FM. Here, $G=0.2\omega_{m}$ and the other parameters are the same as in Fig.~\ref{fig:phonon_num_lim_G}.}
	\label{fig:phonon_num_lim_kappa}
\end{figure}


\begin{eqnarray}\label{e011}
n_{f}=\frac{A_{+}+n_{\mathrm{th}}\gamma}{\Gamma_\mathrm{opt}+\gamma},
\end{eqnarray}
with
\begin{eqnarray}\label{e012}
\Gamma_\mathrm{opt}=A_{-}-A_{+},~~~~A_{-}=\frac{4|G|^2}{\kappa},~~~~A_{+}=\frac{|G|^{2}\kappa}{4\omega_{m}^{2}+\frac{\kappa^{2}}{4}},
\end{eqnarray}
where $A_{-}~(A_{+})$ is the emitting (absorbing) rate of phonon and $\Gamma_\mathrm{opt}$ is the cooling rate. Different from $n_{f}$, which requires very weak optomechanical coupling strength (see Fig.~\ref{fig:phonon_num_lim_G}), the expression of Eq.~(\ref{e010}) is satisfied in the whole stable region $(|G|^{2}<\omega_{m}^{2}/4+\kappa^{2}/16)$, which can be derived via Routh-Hurwitz criterion~\cite{PhysRevA.35.5288}. In order to clearly compare the two steady-state final MPNs with the dynamical results, we numerically solve the differential equations (Eq.~(\ref{e09})) in the absence of FM and show those results in Fig.~\ref{fig:phonon_num_lim_G}. Here, for the dynamical results, we have assumed that the MPN of initial system equals to the average phonon number in thermal equilibrium bath $\langle\delta b^{\dagger}\delta b\rangle(t=0)=n_{\mathrm{th}}=10^{3}$, and the other second order moments are zero. The red solid line is the analytical steady-state final MPN which is given by Eq.~(\ref{e010}), the black dashed line is the exactly numerical result which is obtained by solving Eq.~(\ref{e09}) for the steady-state system, and the blue dot line is the conventional steady-state final MPN which is only satisfied with the very weak optomechanical coupling strength, as shown in Fig.~\ref{fig:phonon_num_lim_G}. In addition, we also give the relationship between the steady-state final MPNs and cavity decay rate, as shown in Fig.~\ref{fig:phonon_num_lim_kappa}. The results indicate that the ground-state cooling of mechanical resonator cannot be achieved when the cavity decay rate is too large or too small in the standard OMSs, which is consistent with previous studies.

In the whole stable region, we give a more general result of the cooling limit in the standard OMSs without FM, which can describe the mechanical cooling better in the broad parameter region. Next, we study the phonon number via solving the differential equations in the presence of FM, which will indicate that the cooling limit can be broken due to the existence of FM.

\subsection{Weak coupling}
\begin{figure}
	\centering
	\includegraphics[width=0.6\linewidth]{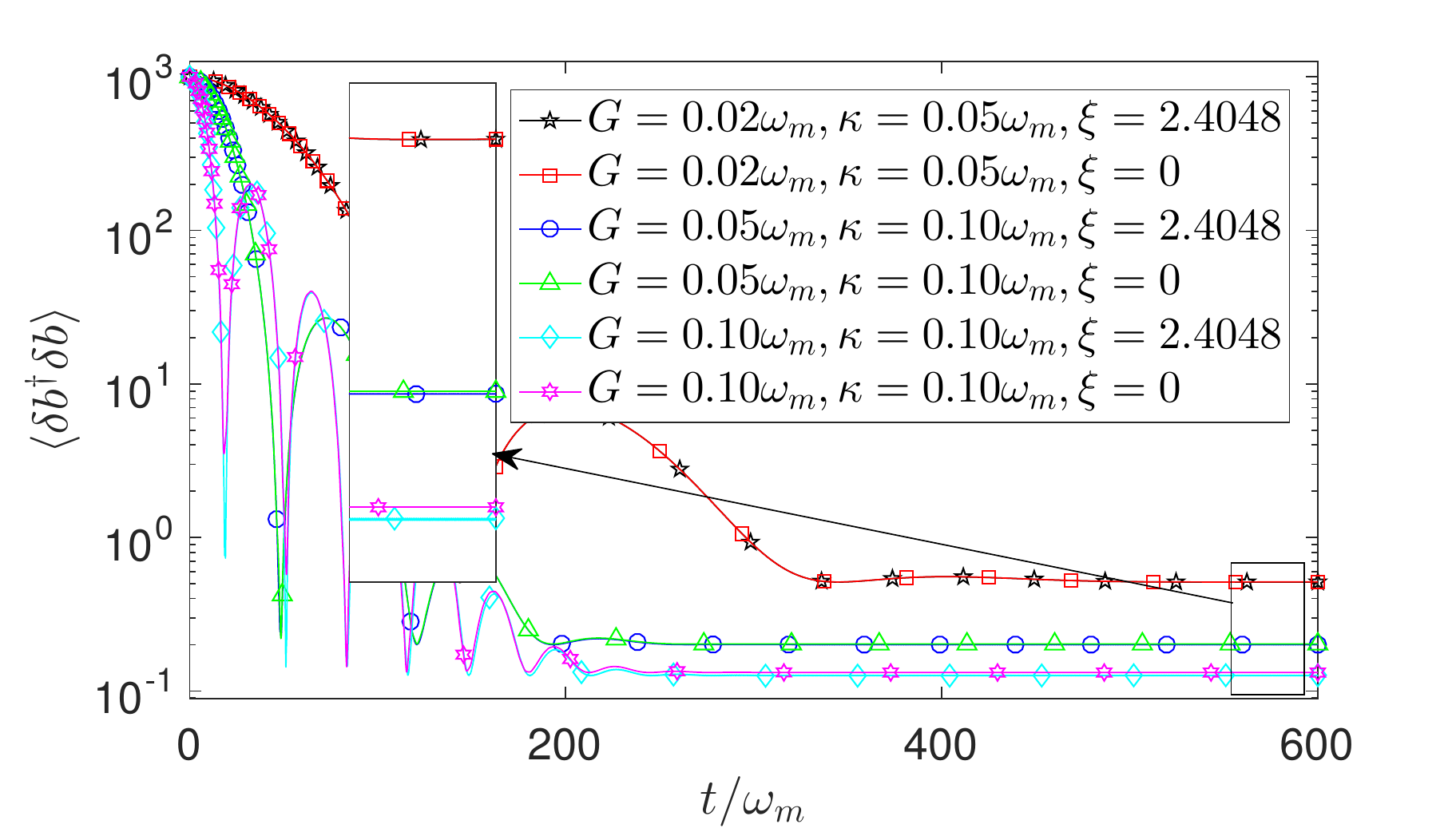}
	\caption{(Color online) The time evolution of MPN $\langle\delta b^{\dagger}\delta b\rangle$ by solving Eq.~(\ref{e09}) for different system parameters with or without FM. The other parameters are chosen as $\Delta_{c}^{'}=\omega_{m}$, $\gamma=10^{-5}\omega_{m}$, $n_{\mathrm{th}}=10^{3}$, and $\nu=5\omega_{m}$.}
	\label{fig:phonon_num_weak}
\end{figure}

For the WC regime, the ground-state cooling of mechanical resonator has been studied extensively in the RSB regime~\cite{PhysRevB.69.125339,PhysRevB.79.193407,PhysRevB.80.144508,PhysRevA.83.013816}. When the decay rate of the cavity is small enough $(\kappa\ll\omega_{m})$, the Stokes heating processes can be suppressed well in the usual red side-band cooling regime without FM. Therefore, the further suppression effect is unobvious via FM, namely, the improving rate of introducing the FM is very small. The result can be verified by showing the dynamical evolution of MPN in Fig.~\ref{fig:phonon_num_weak}. One can see that there is not obvious difference between modulation and no modulation for WC and small cavity decay rate (see Table~\ref{t01}). However, for larger coupling strength $G=0.1\omega_{m}$ and cavity decay $\kappa=0.1\omega_{m}$, the MPN with FM is lower obviously than that without FM (see the cyan diamond line and magenta hexagram line in Fig.~\ref{fig:phonon_num_weak}). This is due to the fact that the Stokes heating processes cannot be suppressed fully for larger coupling strength and cavity decay in the absence of FM.
\begin{table}
	\centering\caption{The steady-state final MPN and the improving rate for different system parameters with or without FM. The other parameters are chosen as $\Delta_{c}^{'}=\omega_{m}$, $\gamma=10^{-5}\omega_{m}$, $n_{\mathrm{th}}=10^{3}$, and $\nu=30\omega_{m}$ when the system is in WC, strong coupling (SC), USC, and USB regimes.}\label{t01}
	\begin{tabular}{|c|c|c|c|c|c|}
		\hline
		~~~~~~~~~ & ~~~\multirow{2}*{$G/\omega_{m}$}~~~ & ~~~\multirow{2}*{$\kappa/\omega_{m}$}~~~ & \multicolumn{2}{c|}{~~~steady-state final MPN~~~} & ~~~\multirow{2}*{improving rate}~~~ \\ 
		\cline{4-5}
		&  &  & ~~without FM~~ & with FM &  \\ 
		\hline 
		\multirow{2}*{WC} 
		& 0.02 & 0.05 & 0.5128 & 0.5123 & 0.98\textperthousand \\ 
		\cline{2-6} 
		& 0.1 & 0.1 & 0.1319 & 0.125 & 5.23\% \\
		\hline
		\multirow{4}*{SC} 
		& 0.3 & 0.1 & 0.1876 & 0.1032 & 44.99\% \\ 
		\cline{2-6}
		& 0.5 & 0.2 & 13.9419 & 0.0527 & 99.62\% \\
		\cline{2-6}
		& 0.6 & 0.3 & fail (diverge) & 0.0363 & success \\ 
		\cline{2-6}
		& 0.9 & 0.5 & fail (diverge) & 0.0233 & success \\
		\hline
		\multirow{2}*{USC}
		& 1.2 & 0.5 & fail (diverge) & 0.0239 & success \\
		\cline{2-6}
		& 1.5 & 0.5 & fail (diverge) & 0.0253 & success \\
		\hline 	
		\multirow{2}*{USB}
		& 0.2 & 4 & 1.5225 & 0.2565 & 83.15\% \\
		\cline{2-6}
		& 0.5 & 10 & 7.1013 & 0.1212 & 98.29\% \\
		\hline 
	\end{tabular} 
\end{table}

In a word, the further cooling from FM is unobvious compared with the usual side-band cooling for the weaker optomechanical coupling in RSB regime. Based on the above analysis, we can easily infer that the cooling approach of utilizing FM to suppress Stokes scattering will stand out with the increment of optomechanical coupling strength and decay rate.

\subsection{Strong coupling}
In the SC regime $(|G|>\kappa)$, the improving rate becomes obvious no matter the decay rate of cavity is large or small. This is very easy to understand based on the above analysis. It is worth noting that the standard OMSs without FM exist the dynamical stability condition $(|G|^{2}<\omega_{m}^{2}/4+\kappa^{2}/16)$. Therefore, the SC region can be roughly divided to the stable $(|G|<0.5\omega_{m})$ and unstable $(|G|>0.5\omega_{m})$ regions, where the MPN of standard OMSs will converge or diverge corresponding to different regions in the absence of FM. Here, we should point out that the OMS with FM can be stable even in the unstable region mentioned above, which can be determined by observing the dynamical behaviors of MPN. Next, we study the mechanical resonator cooling in two different coupling regions, respectively.

\begin{figure}
	\centering
	\includegraphics[width=0.6\linewidth]{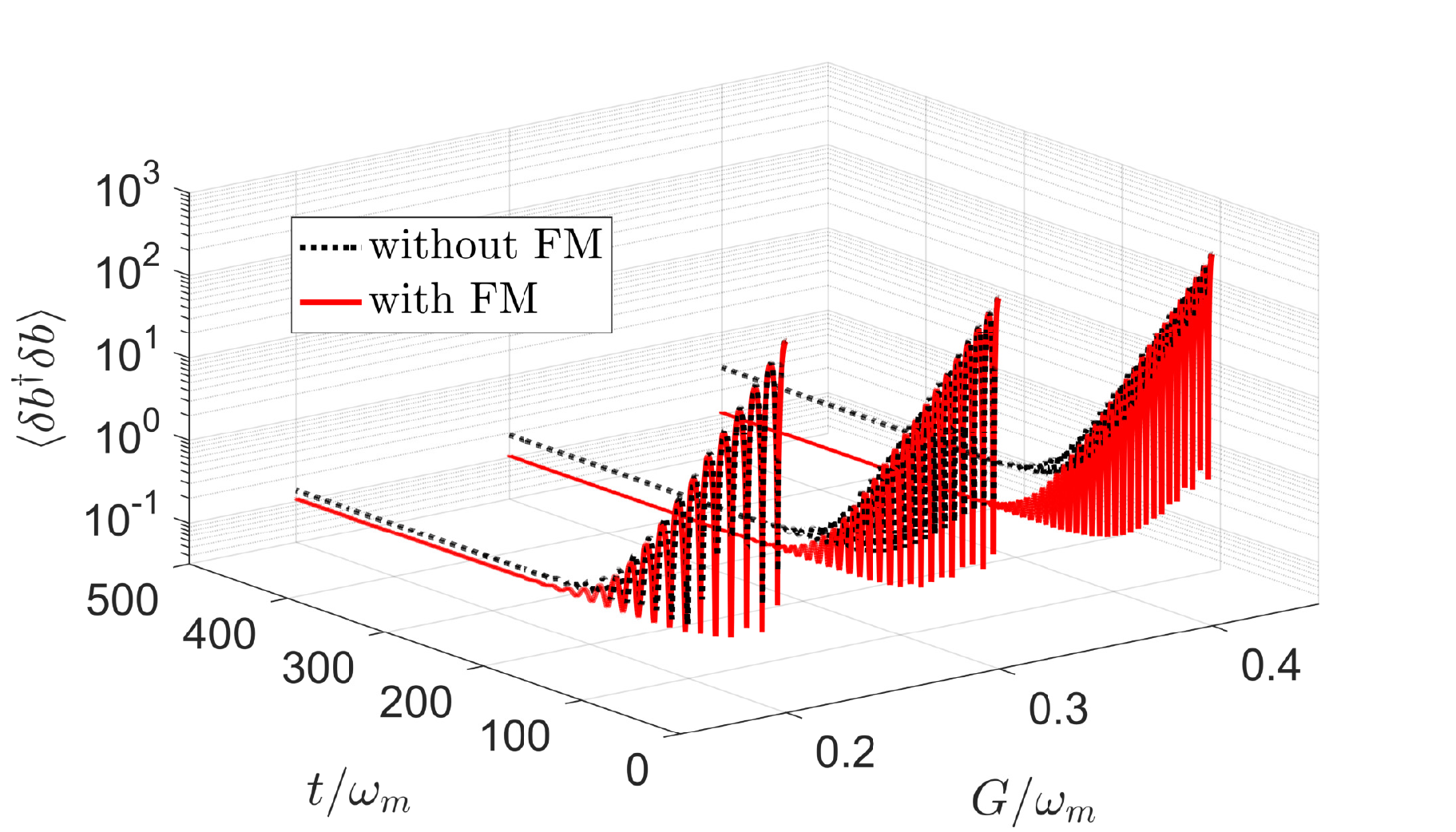}
	\caption{(Color online) The time evolution of MPN $\langle\delta b^{\dagger}\delta b\rangle$ corresponding to different optomechanical coupling strengths with or without FM. Here, $\kappa=0.1\omega_{m}$, $\nu=10\omega_{m}$, and the other parameters are the same as in Fig.~\ref{fig:phonon_num_weak}.}
	\label{fig:phonon_num3D_G_t}
\end{figure}

In the stable region of the standard OMSs, we show the dynamical evolution of MPN corresponding to different optomechanical coupling strengths in Fig.~\ref{fig:phonon_num3D_G_t}, when the FM exists or not. One can see from Fig.~\ref{fig:phonon_num3D_G_t} that the different values of the final MPNs with and without FM increase with increasing the coupling strength. The results indicate that the advantages of cooling with FM are becoming more and more obvious as the system enters the SC regime (see Table~\ref{t01}). We can also see that the final MPN without FM increases with the increase of coupling strength. One reason is the saturation effect of the cooling rate~\cite{PhysRevA.80.063819} and the another is the enhanced Stokes processes. However, the final MPN with FM almost does not change with the increase of coupling strength. This is because the small decay rate of the optical cavity limits the final mechanical resonator cooling in the SC regime.

\begin{figure}
	\centering
	\includegraphics[width=0.6\linewidth]{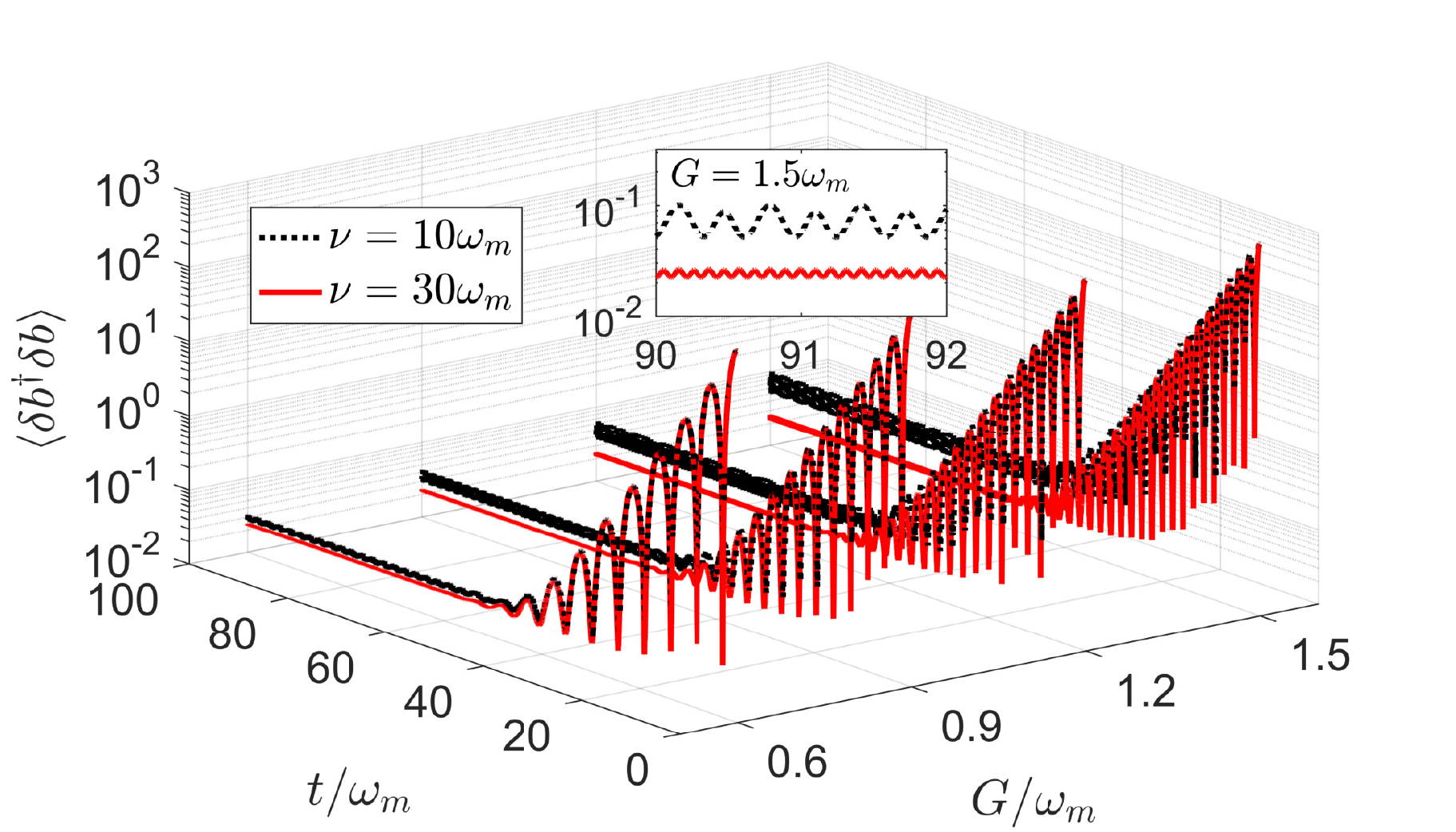}
	\caption{(Color online) The time evolution of MPN $\langle\delta b^{\dagger}\delta b\rangle$ corresponding to different optomechanical coupling strengths with different modulation frequencies. Here, $\kappa=0.5\omega_{m}$ and the other parameters are the same as in Fig.~\ref{fig:phonon_num_weak}.}
	\label{fig:phonon_num3D_G_large}
\end{figure}

In the unstable region of the standard OMSs, where the ground-state cooling of mechanical resonator cannot be achieved due to the divergent behavior of phonon number (see $G>0.5\omega_{m}$ in Table~\ref{t01}), we just show the dynamical evolution of MPN in the presence of FM corresponding to different optomechanical coupling strengths and modulation frequencies, as shown in Fig.~\ref{fig:phonon_num3D_G_large}. The results show that the system is stable due to the existence of FM even in the conventional unstable region. With the increase of modulation frequency, the system becomes more stable and the final MPN is smaller. The fundamental reason is that the large modulation frequency can suppress the Stokes heating processes efficiently. Furthermore, we also note that, for the small modulation frequency in Fig.~\ref{fig:phonon_num3D_G_large}, the final MPN increases with the increase of optomehcanical coupling strength. This is because the small modulation frequency cannot suppress the Stokes heating processes fully when the coupling strength is very large. Moreover, we also show the dynamical evolution of MPN in the USC regime $(|G|>\omega_{m})$, as shown in Fig.~\ref{fig:phonon_num3D_G_large}. We find that the ground-state cooling can also be achieved in USC regime with FM. So we can infer that, for more larger coupling strength, more larger modulation frequency will be needed to achieve the better ground-state cooling of the mechanical resonator.

\subsection{Unresolved side-band regime}
\begin{figure}
	\centering
	\includegraphics[width=0.6\linewidth]{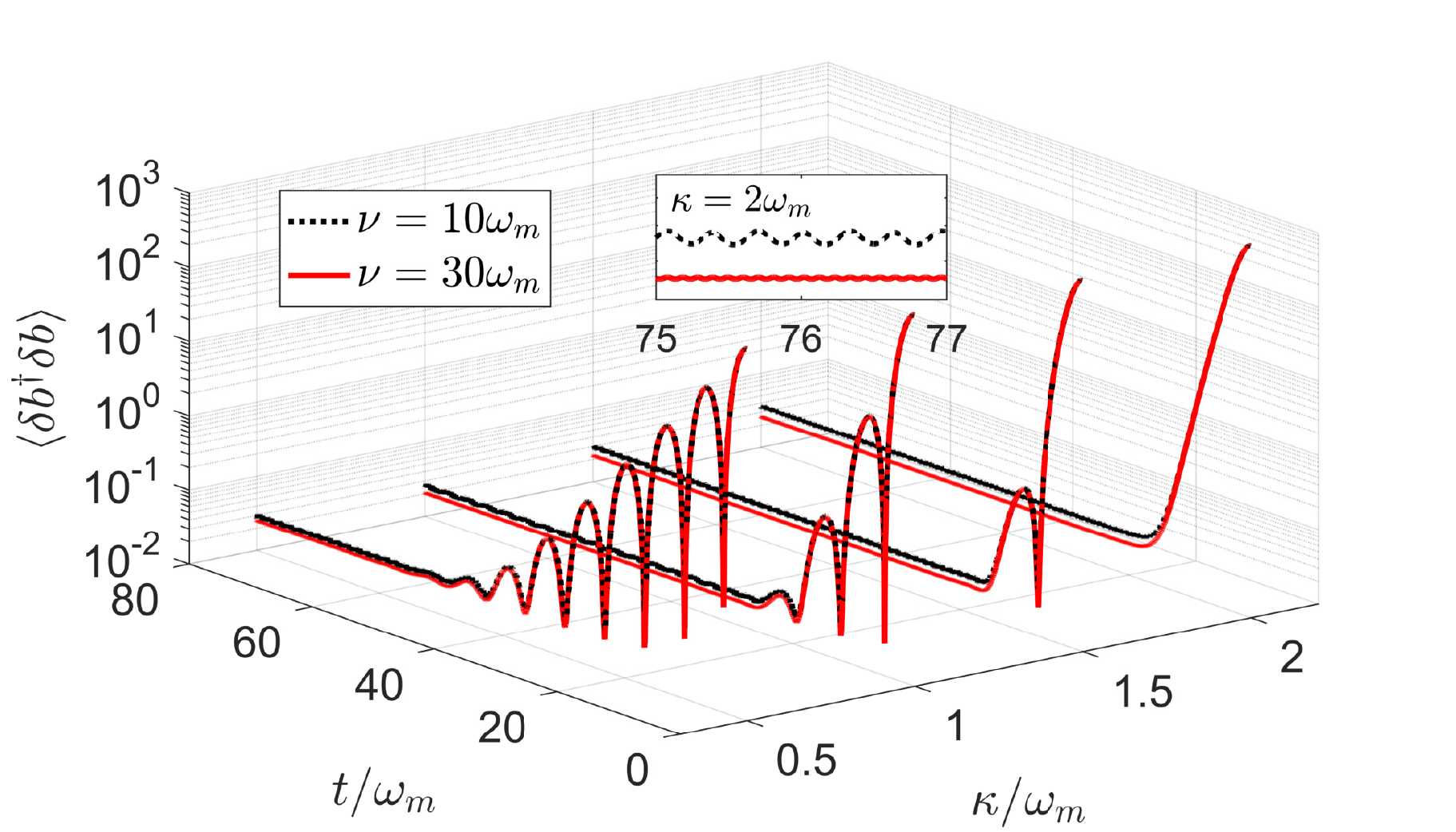}
	\caption{(Color online) The time evolution of MPN $\langle\delta b^{\dagger}\delta b\rangle$ for different cavity decay rates and modulation frequencies. Here, $G=0.2\omega_{m}$, $\nu=10\omega_{m}$, and the other parameters are the same as in Fig.~\ref{fig:phonon_num_weak}.}
	\label{fig:phonon_num3D_kappa_t}
\end{figure}

For the USB regime $(\kappa>\omega_{m})$, the Stokes processes cannot be suppressed well in the standard OMSs, which results in an unsatisfactory ground-state cooling of mechanical resonator (see Table~\ref{t01}). However, in the presence of FM, the nearest resonant Stokes process can be fully suppressed and the other Stokes processes can also be neglected when the modulation frequency is large enough. Similarly, we show the dynamical evolution of MPN corresponding to different cavity decay rates in Fig.~\ref{fig:phonon_num3D_kappa_t}. We find that, in the absence of FM, the ground-state cooling cannot be achieved when the cavity decay rate is too large. However, in the presence of FM, the ground-state cooling can be achieved very well even in the USB regime.

\begin{figure}
	\centering
	\includegraphics[width=0.6\linewidth]{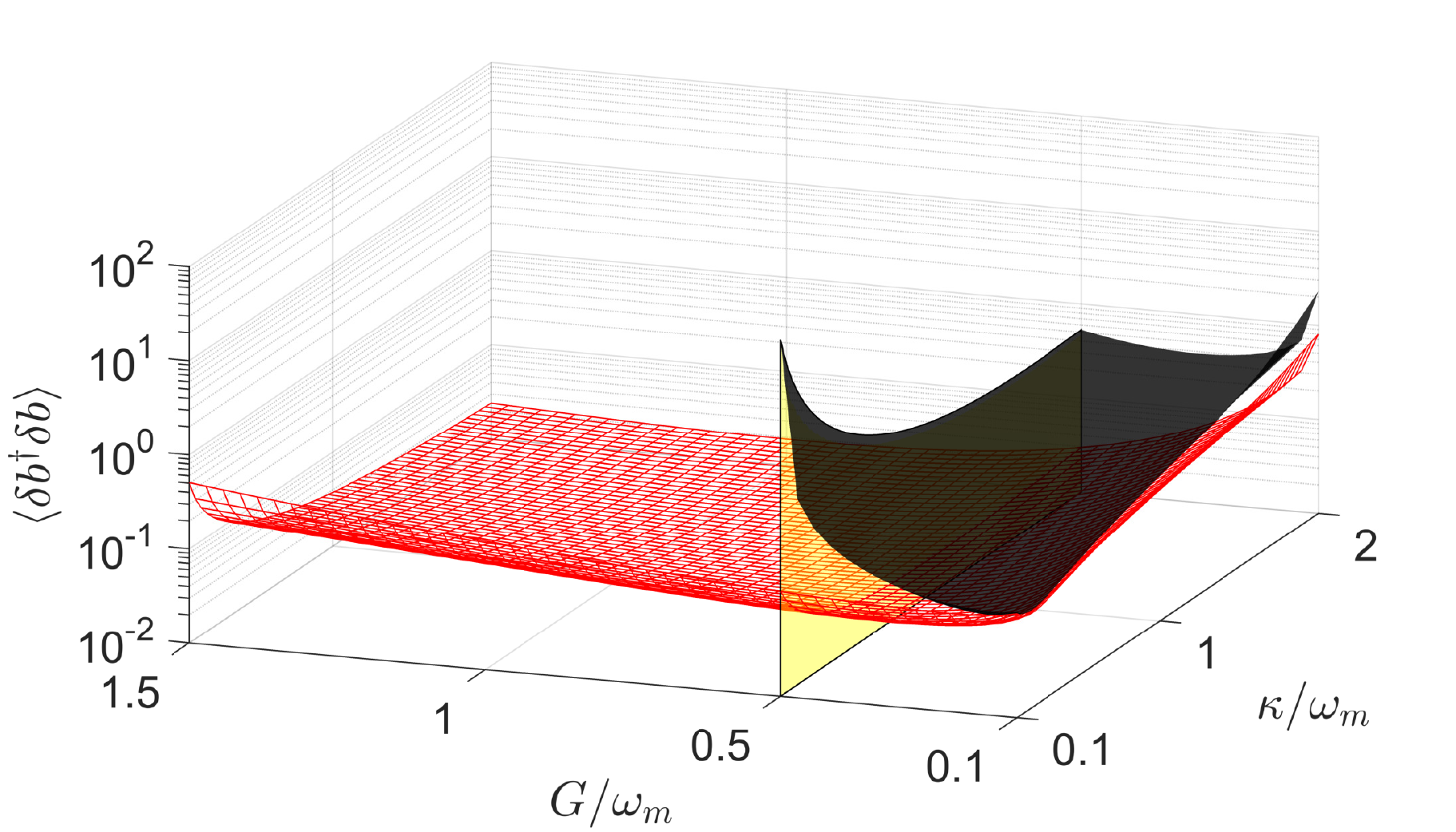}
	\caption{(Color online) The final stable MPN $\langle\delta b^{\dagger}\delta b\rangle$ in relation to the optomechanical coupling strength and cavity decay rate. Here, $\nu=30\omega_{m}$ and the other parameters are the same as in Fig.~\ref{fig:phonon_num_weak}. The black smooth surface represents the final MPN without FM, the red mesh surface represents the final MPN with FM, and the yellow vertical plane is the roughly stable boundary of the standard OMSs without FM.}
	\label{fig:phonon_num_G_kappa}
\end{figure}

In the above discussions, the study of the MPN dynamical evolution is only related to one of the system parameters, i.e., the optomechanical coupling strength or cavity decay rate. To clearly demonstrate the results of their synergistic interaction, we plot the average value of the final MPN after the system is stable, as shown in Fig.~\ref{fig:phonon_num_G_kappa}, where the final stable MPN changes with the optomechanical coupling strength and cavity decay rate, with or without the FM. In the presence of FM, the final MPN of the mechanical resonator is lower than that without FM in the stable region. In addition, we can also achieve the ground-state cooling of mechanical resonator even in the conventional unstable region, as shown the left side of the yellow plane in Fig.~\ref{fig:phonon_num_G_kappa}. The results indicate that the system with FM is still stable and the ground-state cooling can also be achieved even in the conventional unstable region for the standard OMSs.

\subsection{Small modulation frequency}
\begin{figure}
	\centering
	\includegraphics[width=0.6\linewidth]{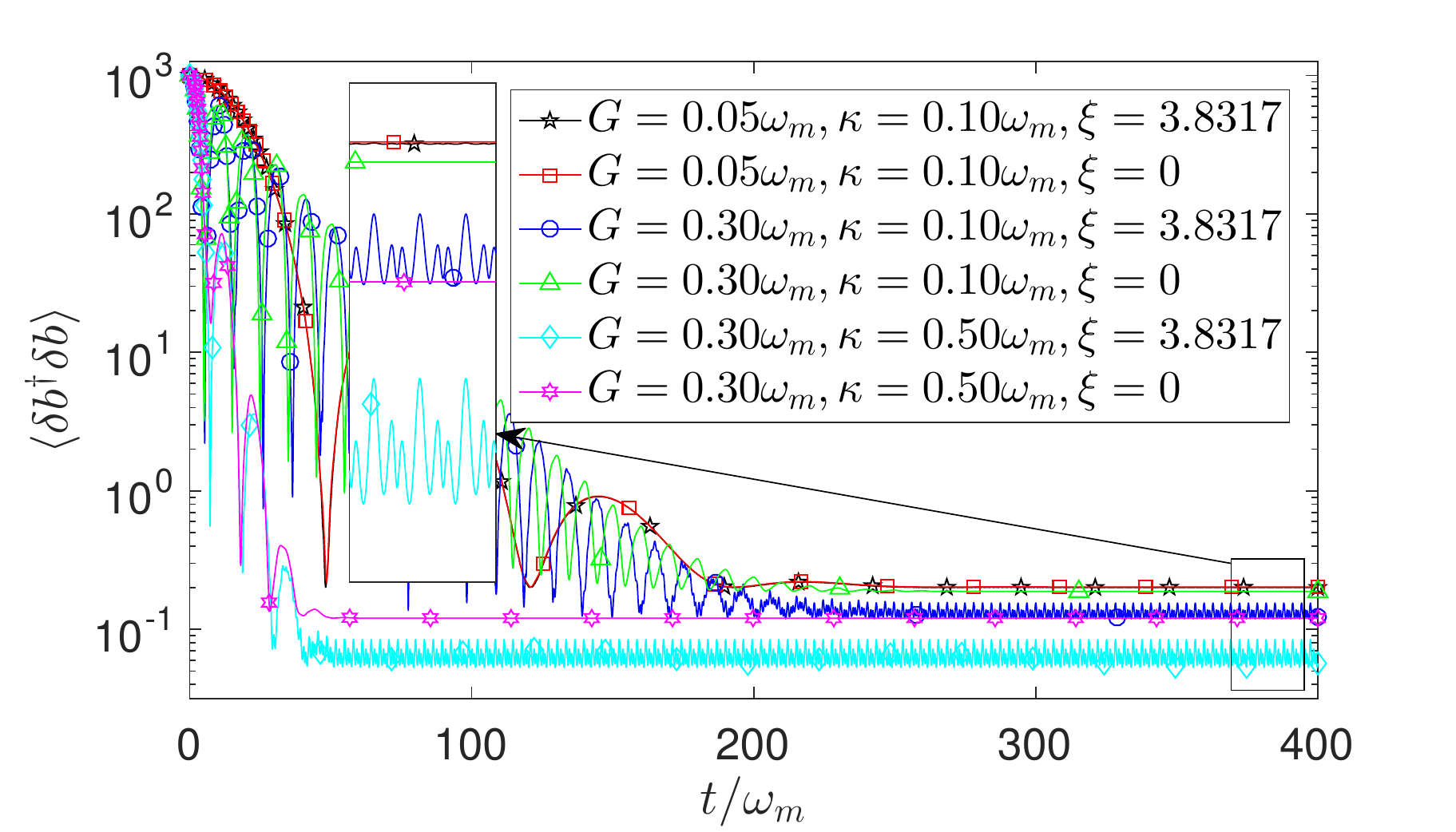}
	\caption{(Color online) The time evolution of MPN $\langle\delta b^{\dagger}\delta b\rangle$ corresponding to different system parameters with or without the modulation. Here $\nu=2\omega_{m}$, $k_{0}=-1$, and the other parameters are the same as in Fig.~\ref{fig:phonon_num_weak}.}
	\label{fig:phonon_number_small_modu}
\end{figure}

In the above subsections, in order to suppress the Stokes heating processes better, the large modulation frequency is necessary. In practice, however, too large modulation frequency may not be conducive to experimental implementation, although various of schemes about the FM of mechanical resonator have been reported theoretically and experimentally~\cite{PhysRevA.91.023818,JAP.1.2785018,nnano.2013.232,s151026478,nnano.2014.168,nl500879k,1367-2630-9-2-035}. Therefore, we consider the effect of a small modulation frequency on the system in this subsection. We set the modulation frequency as $\nu=2\omega_{m}$, where one of the Stokes processes $(k_{0}=-1)$ can be resonant ideally like the anti-Stokes process in the red detuning side-band resonant regime. Here, the resonant Stokes process can also be suppressed by making coupling strength zero, for example, with the choice of $\xi=3.8317$, the effective optomechanical coupling strength $GJ_{-1}(3.8317)=0$. However, the other Stokes processes $(k\neq-1)$ cannot be suppressed well due to $J_{k\neq-1}(3.8317)\neq0$ at this time. The dynamical evolution of MPN with the small modulation frequency is shown in Fig.~\ref{fig:phonon_number_small_modu} for different system parameters. The results indicate that the cooling effect resorting to FM is still better than the standard OMSs without FM under these system parameters. The influences of MPN by changing system parameters are also similar to the situation of large modulation frequency.

\subsection{Asynchronous frequency modulation}
In previous discussions, we have assumed that the modulations of optical and mechanical components are identical and synchronous to demonstrate clearly the physical mechanism of the further cooling via FM. Here, we will study the effects of asynchronous FM, e.g., different amplitudes, phases, and frequencies of the modulation. The Hamiltonian of the modulation parts changes to $H_{M}=[\xi_{1}\nu_{1}\cos(\nu_{1}t)a^{\dagger}a+\xi_{2}\nu_{2}\cos(\nu_{2}t+\theta)b^{\dagger}b]/2$, where $\xi_{1}(\xi_{2})$ and $\nu_{1}(\nu_{2})$ are normalized amplitude and frequency of the optical (mechanical) modulation, respectively, and $\theta$ is the relative phase of two modulations.

\begin{figure}
	\centering
	\includegraphics[width=0.6\linewidth]{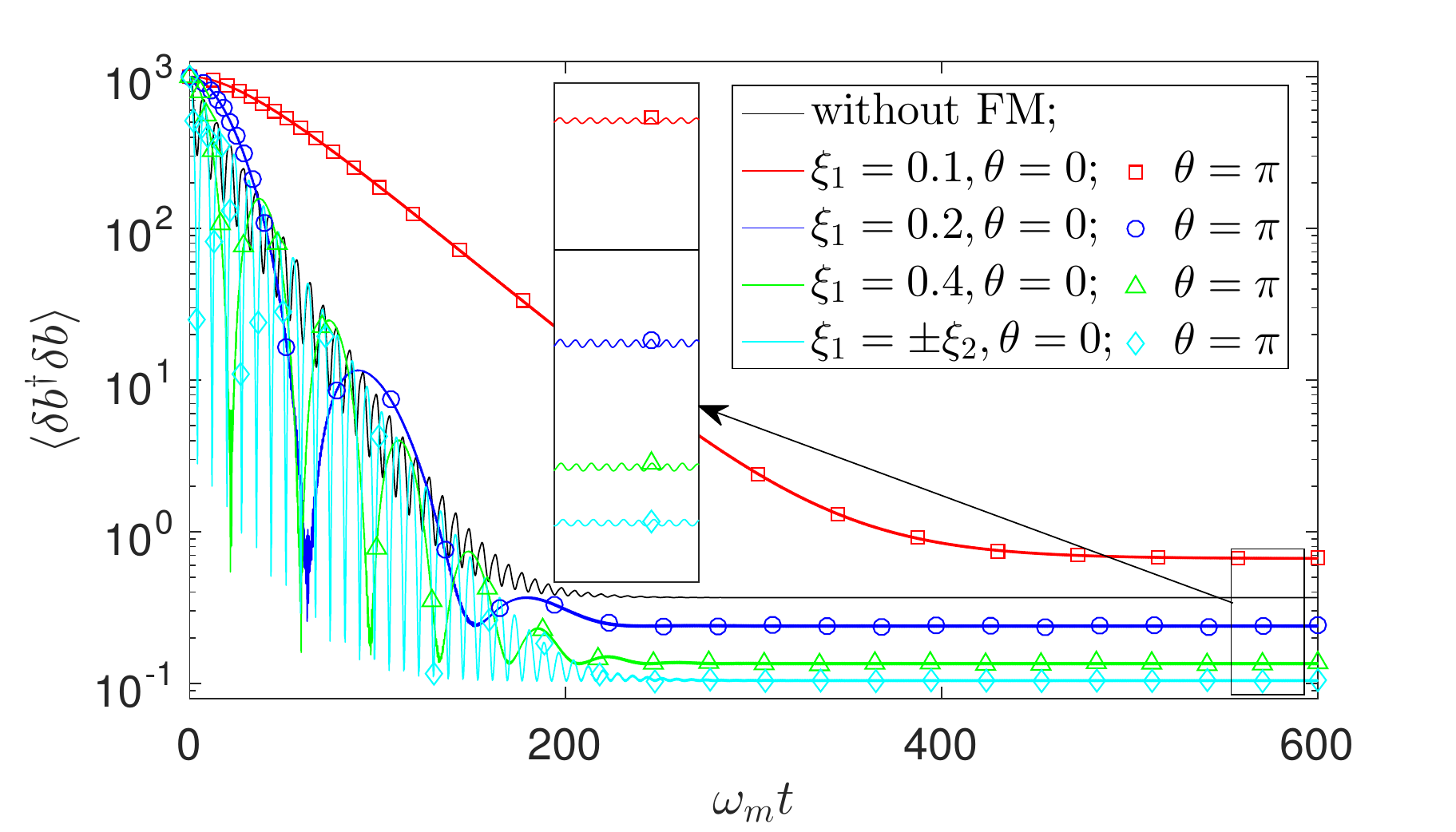}
	\caption{(Color online) The time evolution of MPN $\langle\delta b^{\dagger}\delta b\rangle$ corresponding to different amplitudes of the FM. Here $G=0.4\omega_{m}$, $\kappa=0.1\omega_{m}$, $\nu_{1}=\nu_{2}=10\omega_{m}$, $\xi_{2}=2\xi_{0}-\xi_{1}(\xi_{2}=\xi_{1}-2\xi_{0})$ for $\theta=0(\theta=\pi)$, and the other parameters are the same as in Fig.~\ref{fig:phonon_num_weak}.}
	\label{fig:different_amp}
\end{figure}

Firstly, we study the effects of different optical and mechanical modulation amplitudes, where the other modulation parameters are identical, i.e., $\xi_{1}\neq\xi_{2}$, $\theta=0$, and $\nu_{1}=\nu_{2}$. The interaction Hamiltonian can be written as
\begin{eqnarray}\label{e013}
\widetilde{H}_{\xi}&=&\sum_{k=-\infty}^{\infty}\Bigg[-GJ_{k}\left(\frac{\xi_{1}-\xi_{2}}{2}\right)\delta a^{\dagger}\delta be^{i(\Delta_{c}^{'}-\omega_{m}+k\nu)t}\cr\cr
&&-GJ_{k}\left(\frac{\xi_{1}+\xi_{2}}{2}\right)\delta a^{\dagger}\delta b^{\dagger}e^{i(\Delta_{c}^{'}+\omega_{m}+k\nu)t}+\mathrm{H.c.}\Bigg].
\end{eqnarray}
Based on the previous analyses, the mechanical cooling can be improved if the nearest resonant Stokes heating process is suppressed with $(\xi_{1}+\xi_{2})/2=\xi_{0}$, where $J_{k_{0}}(\xi_{0})=0$. It is worth noting that the beam-splitter interactions are also related to the different modulation amplitudes. Therefore, the improving mechanical cooling cannot be achieved if the difference value $(\xi_{1}-\xi_{2})/2$ is close to $\xi_{0}$, which greatly reduces the optomechanical coupling. In addition, the similar results can also be obtained with $\theta=\pi$, where the beam-splitter and two-mode squeezing interactions exchange their coupling strengths. The dynamical results of MPN are shown in Fig.~\ref{fig:different_amp} with different amplitudes.

\begin{figure}
	\centering
	\includegraphics[width=0.6\linewidth]{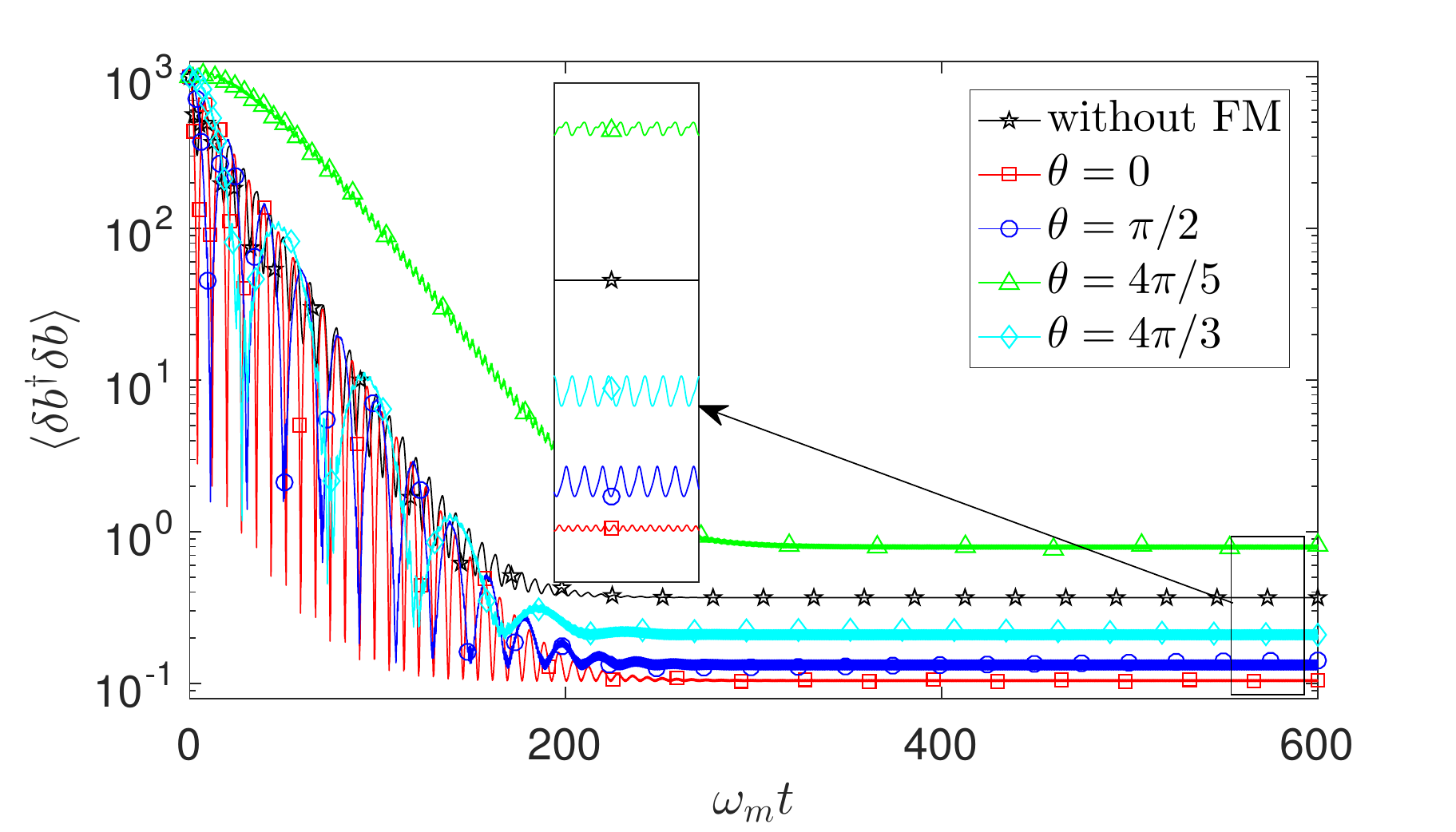}
	\caption{(Color online) The time evolution of MPN $\langle\delta b^{\dagger}\delta b\rangle$ corresponding to different phases of the FM. Here $\xi_{1}=\xi_{2}=\xi_{0}$ and the other parameters are the same as in Fig.~\ref{fig:different_amp}.}
	\label{fig:different_pha}
\end{figure}

For the different phases of FM, we also study the effects by changing $\theta$ and show the results in Fig.~\ref{fig:different_pha}. We find that the improving mechanical cooling can also be achieved when $\theta$ is far from $\pi$ with $\xi_{1}=\xi_{2}$ and $\nu_{1}=\nu_{2}$. That is easy to understand due to the exchange of coupling strengths for Eq.~(\ref{e013}) when the relative phase $\theta$ equals to $\pi$, where the coupling strength of the nearest resonant two-mode squeezing interaction is $GJ_{0}(0)=G$. And the Stokes processes cannot be suppressed so that the improving mechanical cooling is not achieved.

\begin{figure}
	\centering
	\includegraphics[width=0.6\linewidth]{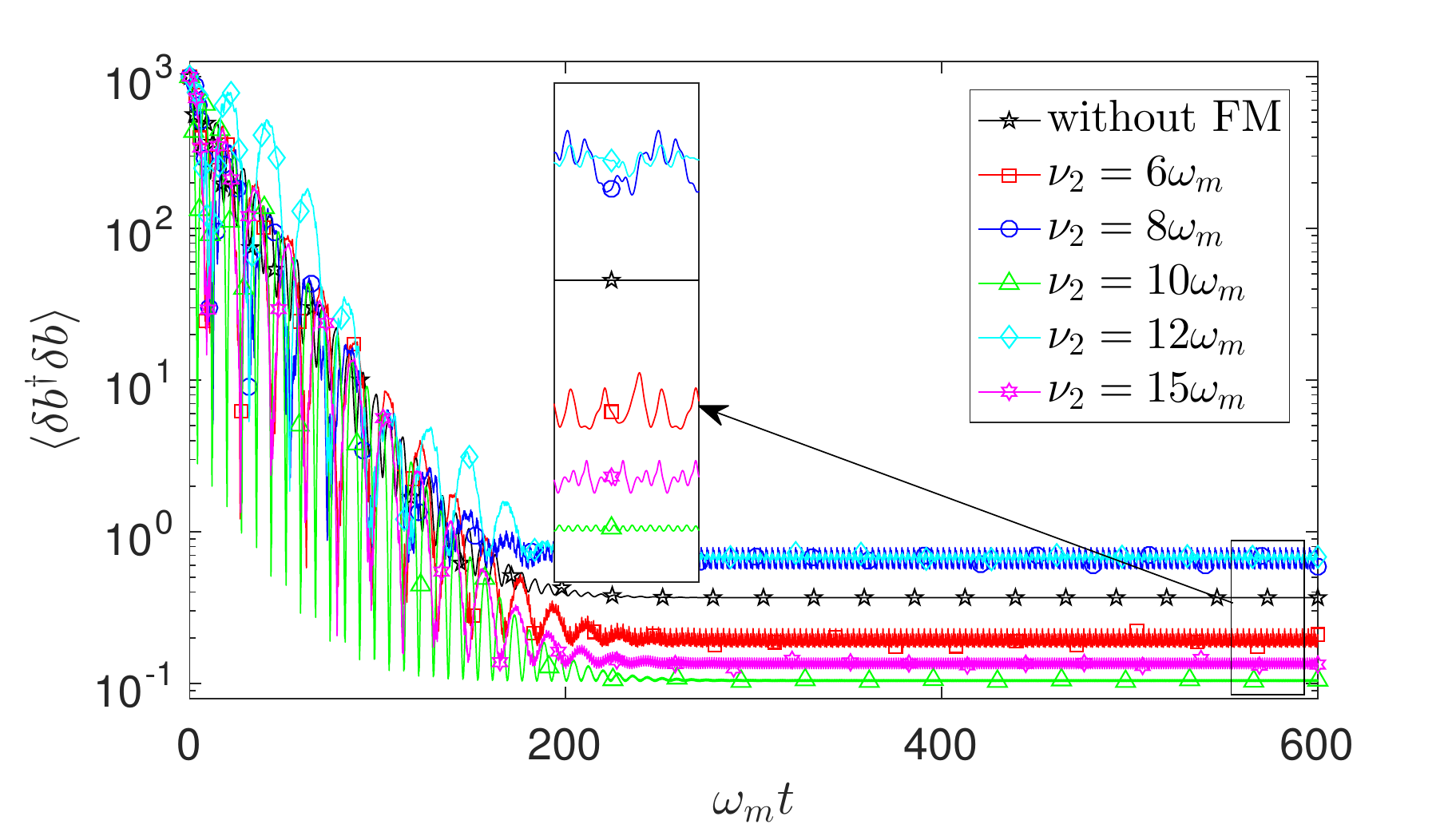}
	\caption{(Color online) The time evolution of MPN $\langle\delta b^{\dagger}\delta b\rangle$ corresponding to different frequencies of the FM. Here $\theta=0$, $\nu_{1}=10\omega_{m}$, and the other parameters are the same as in Fig.~\ref{fig:different_amp}.}
	\label{fig:different_fre}
\end{figure}

Lastly, we study the effects of different modulation frequencies, i.e., $\nu_{1}\neq\nu_{2}$. We find that the improving mechanical cooling cannot be achieved when the two modulation frequencies satisfy $|\nu_{1}-\nu_{2}|\simeq2\omega_{m}$, as shown in Fig.~\ref{fig:different_fre}. However, the steady-state final MPN with FM is still lower than that in the conventional OMS without FM when the difference value of the two modulation frequencies is far from $2\omega_{m}$. The foremost reason is that the detuning of the Stokes processes is decreased and results in the suppressed effect reducing. Furthermore, even if all the  modulation parameters are different, e.g., $\xi_{1}=1$, $\xi_{2}=3.8096$, $\theta=\pi/4$, $\nu_{1}=10\omega_{m}$, and $\nu_{2}=15\omega_{m}$, the average steady-state final MPN is $\langle\delta b^{\dagger}\delta b\rangle=0.1327$, which is still lower than the result of conventional OMS without FM $\langle\delta b^{\dagger}\delta b\rangle=0.3677$.

\section{Conclusions}\label{sec.4}
In conclusion, we have proposed a proposal for improving the cooling of micromechanical resonator in OMS through introducing FM method. We proposed a deeper insight of the optomechanical cooling when the frequencies of cavity and mechanical resonator are modulated periodically, which would provide a guide for optomechanical cooling experiments to cool the mechanical resonator below the QBL. Here, we have redefined a more general QBL in the standard OMSs without FM, which is satisfied in the whole stable region. In the presence of FM, we analyze and demonstrate the reasons of cooling mechanical resonator to the lower MPN, which are due to the fact that the Stokes heating processes can be suppressed fully such that the QBL of mechanical cooling can be broken via the FM. In addition, we also study the dynamical evolution of MPN by solving the differential equations of second order moments numerically, which are derived strictly from the quantum master equation. We find that the lower MPN can be reached when the modulation frequency is large enough, even if the coupling strength ranges from WC to USC regimes and the cavity decay rate ranges from RSB to USB regimes. And the improving rate of mechanical cooling with FM becomes more and more larger with the increase of optomechanical coupling strength or cavity decay rate. However, in the standard OMSs without FM, the ground-state cooling of mechanical resonator cannot be achieved when either the coupling strength or cavity decay rate is too large. The results show that, in the presence of FM, the mechanical ground-state cooling is not limited to the conventional stability boundary and the RSB regime, and in the meanwhile the proposed proposal is still feasible even with more board system parameters. Moreover, we also give the mechanical cooling results when the modulation frequency is not large enough, where the Stokes processes cannot be suppressed well. At last, we discuss respectively the effects of asynchronous FM and find that the improving mechanical can be achieved even with different modulation. Our proposal would open up the possibility for cooling the mechanical resonator beyond the QBL of the standard optomechanical cooling.

\begin{center}
{\bf{ACKNOWLEDGMENTS}}
\end{center}
This work was supported by the National Natural Science Foundation of China under Grant Nos. 11465020, 11264042, 61465013, 61575055, and the Project of Jilin Science and Technology Development for Leading Talent of Science and Technology Innovation in Middle and Young and Team Project under Grant No. 20160519022JH.

\appendix
\section{Linearizing the system Hamiltonian in the presence of frequency modulation}\label{App1}
In the presence of FM, the system Hamiltonian of the standard cavity OMS reads
\begin{eqnarray}\label{Ae01}
H=\omega_{c}a^{\dagger}a+\omega_{m}b^{\dagger}b-ga^{\dagger}a(b^{\dagger}+b)+(Ea^{\dagger}e^{-i\omega_{l}t}+E^{\ast}ae^{i\omega_{l}t})
	+\frac{1}{2}\xi\nu\cos(\nu t)(a^{\dagger}a+b^{\dagger}b),
\end{eqnarray}
where $a~(b)$ and $a^{\dagger}~(b^{\dagger})$ represent the annihilation and creation operators for optical (mechanical) mode with the corresponding frequency $\omega_{c}~(\omega_{m})$, respectively. The parameter $g$ is the single photon optomechanical coupling rate. The parameters $E$ and $\omega_{l}$ are the driving amplitude and frequency, respectively. By performing a rotating transformation defined by $V_{1}=\exp[-i\omega_{l}ta^{\dagger}a]$, the transformed Hamiltonian $\widetilde{H}=V_{1}^{\dag}HV_{1}-iV_{1}^{\dag}\dot{V}_{1}$ becomes
\begin{eqnarray}\label{Ae02}
\widetilde{H}=\Delta_{c}a^{\dagger}a+\omega_{m}b^{\dagger}b-ga^{\dagger}a(b^{\dagger}+b)+(Ea^{\dagger}+E^{\ast}a)
	+\frac{1}{2}\xi\nu\cos(\nu t)(a^{\dagger}a+b^{\dagger}b),
\end{eqnarray}
where $\Delta_{c}=\omega_{c}-\omega_{l}$ is the cavity laser detuning. 

The quantum Langevin equations are given by
\begin{eqnarray}\label{Ae03}
\dot{a}&=&-i\Delta_{c}a-\frac{\kappa}{2}a+iga(b^{\dagger}+b)-iE-i\frac{1}{2}\xi\nu\cos(\nu t)a-\sqrt{\kappa}a_{\mathrm{in}},\cr\cr
\dot{b}&=&-i\omega_{m}b-\frac{\gamma}{2}b+iga^{\dagger}a-i\frac{1}{2}\xi\nu\cos(\nu t)b-\sqrt{\gamma}b_{\mathrm{in}},
\end{eqnarray}
where $\kappa$ and $\gamma$ are the decay rate of optical cavity and the damping rate of mechanical resonator, respectively. $a_{\mathrm{in}}$ and $b_{\mathrm{in}}$ are the corresponding noise operators. Under the strongly coherent laser driving, we can apply a displacement transformation to linearize  Eq.~(\ref{Ae03}), i.e., $a=\alpha+\delta a$ and $b=\beta+\delta b$, where $\alpha$ and $\beta$ are $c$-numbers representing the displacement mean values of the cavity and mechanical resonator modes. $\delta a$ and $\delta b$ are the operators relating to the quantum fluctuations of the cavity and mechanical resonator modes. Equation~(\ref{Ae03}) can be separated to two different sets of equations, one for the mean values, and the other for the fluctuations, which are given by
\begin{eqnarray}\label{Ae04}
\dot{\alpha}&=&-i\Delta_{c}^{'}\alpha-\frac{\kappa}{2}\alpha-iE-i\frac{1}{2}\xi\nu\cos(\nu t)\alpha,\cr\cr
\dot{\beta}&=&-i\omega_{m}\beta-\frac{\gamma}{2}\beta+ig|\alpha|^{2}-i\frac{1}{2}\xi\nu\cos(\nu t)\beta,\cr\cr
\dot{\delta a}&=&-i\Delta_{c}^{'}\delta a-\frac{\kappa}{2}\delta a+iG(\delta b^{\dagger}+\delta b)
			-i\frac{1}{2}\xi\nu\cos(\nu t)\delta a-\sqrt{\kappa}a_{\mathrm{in}},\cr\cr
\dot{\delta b}&=&-i\omega_{m}\delta b-\frac{\gamma}{2}\delta b+iG\delta a^{\dagger}+iG^{\ast}\delta a
			-i\frac{1}{2}\xi\nu\cos(\nu t)\delta b-\sqrt{\gamma}b_{\mathrm{in}},
\end{eqnarray}
where $\Delta_{c}^{'}=\Delta_{c}-g(\beta^{\ast}+\beta)$ is the effective detuning modified by optomechanical coupling, $G=g\alpha$ is linearized optomechanical coupling strength, and we have neglected the nonlinear terms $ig\delta a(\delta b^{\dagger}+\delta b)$ 
and $ig\delta a^{\dagger}\delta a$ due the strong coherent driving conditions. Then we obtain the linearized Hamiltonian (see Eq.~(\ref{e04}) in the main text) in the presence of FM.

\section{The conventional steady-state final mean phonon number}\label{App2}
For the standard OMSs, the MPN can be derived and calculated by using the rate equations of the mechanical resonator, where we need calculate the spectral density of the optical force using the quantum noise approach. Firstly, we give the effective Hamiltonian of the standard optomechanical (see Eq.~(\ref{e02}) in the main text) in the absence of FM,
\begin{eqnarray}\label{Be01}
H_{L}=\Delta_{c}^{'}\delta a^{\dagger}\delta a+\omega_{m}\delta b^{\dagger}\delta b
-(G\delta a^{\dagger}+G^{\ast}\delta a)(\delta b^{\dagger}+\delta b),
\end{eqnarray}
where we can obtain the optical force $F=(G\delta a^{\dagger}+G^{\ast}\delta a)/x_{\mathrm{ZPF}}$. The quantum noise spectrum of the optical force is given by the Fourier transform of the autocorrelation function $S_{FF}(\omega)=\int dte^{i\omega t}\langle F(t)F(0)\rangle$, which is calculated easily in the frequency domain. In the WC regime, the optomechanical coupling can be regarded as a perturbation to the optical field, where the spectral density of the optical force can be calculated properly without considering optomechanical coupling. In the frequency domain, the cavity mode fluctuation operator $\delta a(\omega)$ can be obtained in the  absence of the optomechanical coupling,
\begin{eqnarray}\label{Be02}
\delta a(\omega)=\frac{\sqrt{\kappa}a_{\mathrm{in}}(\omega)}{i(\omega-\Delta_{c}^{'})-\frac{\kappa}{2}}.
\end{eqnarray}

According to the Hamiltonian in Eq.~(\ref{Be01}), we can write the rate equations of the mechanical resonator as
\begin{eqnarray}\label{Be03}
\dot{P}_{n}&=&\Gamma_{n+1\rightarrow n}P_{n+1}+\Gamma_{n-1\rightarrow n}P_{n-1}-\Gamma_{n\rightarrow n-1}P_{n}
-\Gamma_{n\rightarrow n+1}P_{n}\cr\cr
&&+\gamma(n_{\mathrm{th}}+1)(n+1)P_{n+1}+\gamma n_{\mathrm{th}}nP_{n-1}-\gamma(n_{\mathrm{th}}+1)nP_{n}-\gamma n_{\mathrm{th}}(n+1)P_{n},
\end{eqnarray}
where $P_{n}$ is the occupation probability of the mechanical Fock staet $|n\rangle$ with $n$ phonons. $\Gamma_{n\rightarrow m}$ is the transition rate of phononic state from $|n\rangle$ to $|m\rangle$ induced by the optomechanical coupling. Using the Fermi golden rule and the above calculations, we can obtain
\begin{eqnarray}\label{Be04}
\Gamma_{n\rightarrow n-1}&=&nA_{-}=nS_{FF}(\omega_{m})x_{\mathrm{ZPF}}^{2}
=\frac{n|G|^{2}\kappa}{(\omega_{m}-\Delta_{c}^{'})^{2}+\frac{\kappa^{2}}{4}},\cr\cr
\Gamma_{n-1\rightarrow n}&=&nA_{+}=nS_{FF}(-\omega_{m})x_{\mathrm{ZPF}}^{2}
=\frac{n|G|^{2}\kappa}{(-\omega_{m}-\Delta_{c}^{'})^{2}+\frac{\kappa^{2}}{4}},
\end{eqnarray}
where $A_{-}~(A_{+})$ is the emitting (absorbing) rate of phonon. By solving Eq.~(\ref{Be03}) with $\dot{P}_{n}=0$, the steady-state final MPN can be obtained as
\begin{eqnarray}\label{Be05}
n_{f}=\frac{A_{+}+n_{\mathrm{th}}\gamma}{\Gamma_{\mathrm{opt}}+\gamma}.
\end{eqnarray}
In the absence of intrinsic mechanical damping rate $(\gamma=0)$, we also obtain the fundamental quantum limit of cooling
\begin{eqnarray}\label{Be06}
n_{c}=\frac{A_{+}}{A_{-}-A_{+}}.
\end{eqnarray}

\end{document}